\documentclass[structabstract]{aa} 
 
\usepackage{graphicx}
\usepackage{txfonts}
\usepackage{natbib}
\bibpunct{(}{)}{;}{a}{}{,}

\begin{document}

   \title{The delay time distribution of type Ia supernovae: \\
          a comparison between theory and observation
         }

   \author{N. Mennekens\inst{1}
           \and D. Vanbeveren\inst{1,2}
           \and J.P. De Greve\inst{1}
           \and E. De Donder\inst{1,3}
          }

   \institute{Astrophysical Institute, Vrije Universiteit Brussel, Pleinlaan 2, 1050 Brussels, Belgium\\
              \email{nmenneke@vub.ac.be}
              \and Groep T - Leuven Engineering College, K.U.Leuven Association, Andreas Vesaliusstraat 13, 3000 Leuven, Belgium
              \and Belgian Institute for Space Aeronomy (BIRA-IASB), Ringlaan 3, 1180 Brussels, Belgium
             }

   \date{Received 22 January 2010 / Accepted 10 March 2010}

  \abstract
   {}
   {We investigate the contribution of different formation scenarios for type Ia supernovae in elliptical galaxies. The single degenerate scenario (a white dwarf accreting from a late main sequence or red giant companion) is tested against the double degenerate scenario (the spiral-in and merging of two white dwarfs through the emission of gravitational wave radiation).}
   {We use a population number synthesis code incorporating the latest physical results in binary evolution, and allowing to differentiate between certain physical scenarios (e.g. description of common envelope evolution) and evolutionary parameters (e.g. mass transfer efficiency during Roche lobe overflow). The obtained theoretical distributions of the delay times of type Ia supernovae are compared to those which are observed, both in morphological shape and absolute number of events. The critical dependency of these distributions on certain parameters is used to constrain the values of the latter.}
   {We find that the single degenerate scenario alone cannot explain the morphological shape of the observational delay time distribution, while the double degenerate scenario (or a combination of both) can. Most of these double degenerate type Ia supernovae are created through a normal quasi-conservative Roche lobe overflow followed by a common envelope phase, not through two successive common envelope phases. This may cast doubt on the use in other studies of analytical formalisms to determine delay times. In terms of absolute number, theoretical supernova Ia rates in old elliptical galaxies lie a factor of at least three below the observed ones. We propose a solution involving the effect of rotation on the evolution of intermediate mass binaries.}
   {}

            
   \titlerunning{The delay time distribution of type Ia supernovae}
   
   \authorrunning{N. Mennekens et al.}

   \maketitle

\section{Introduction}

Type Ia supernovae (SNe Ia) are among the most powerful events observed in the universe. Their seemingly identical nature has led them to be most useful distance indicators, or standard candles, in cosmology, while their chemical enrichment of the interstellar medium is of paramount importance for the chemical evolution of galaxies. Large amounts of the chemical element iron (Fe) are released during a SN Ia. As an often quoted example, the present mass fraction of this element in the solar neighborhood could not be explained without the inclusion of SNe Ia in chemical evolution models.

The current consensus is that SNe Ia originate from the thermonuclear disruption of a white dwarf (WD) which is no longer able to support its own mass through degenerate electron pressure, as a result of approaching or exceeding the Chandrasekhar limit \citep[see e.g.][]{livio2001}. However, many open questions remain: whether only carbon-oxygen (C-O) WDs are involved, whether the disruption is a detonation or a deflagration, whether the event takes place at the Chandrasekhar mass or at a sub-Chandrasekhar mass, etc. \citep[see e.g.][]{branch1995}. Perhaps most importantly, even the exact progenitor system of a SN Ia has not been unambiguously established. There exist different scenarios, some of them involving one degenerate object accreting from a less evolved companion in a multiple star system, others assuming that SNe Ia are caused by the merger of two WDs. All of these elements can have important consequences for the applications of SNe Ia, especially for their use as standard candles.

In this work, we differentiate between the two most invoked formation scenarios for SNe Ia: the single degenerate (SD) and double degenerate (DD) model. In the SD model \citep[see e.g.][]{nomoto1982} a WD accretes matter through mass transfer from a companion in a close binary system. The companion can be either a late main sequence (MS) star or a red giant (RG). In the DD model \citep[see e.g.][]{iben1984,webbink1984} the SN Ia is caused by the merger of two WDs, the combined mass of which equals or exceeds the Chandrasekhar mass. Their merger is the result of a spiral-in caused by the emission of gravitational wave radiation (GWR). The validity of this scenario has often been questioned, since the exact detonation mechanism is unclear. It has been argued that a merger may instead result in an off-center carbon ignition and the formation of a oxygen-neon-magnesium WD or a neutron star \citep[see e.g.][]{saio1998}. However, by including rotation \citet{piersanti2003} have shown that the accretor star can cool down, causing a greater expansion of its outer layers and a lowering of the critical angular velocity. This limits the mass accretion rate and prevents the occurrence of a gravitational instability, allowing mass accumulation up to the explosive central ignition of carbon, leading to a SN Ia.

\subsection{The delay time distribution}

The question that will be addressed in this work is which of these two scenarios is the main contributor to the SN Ia explosions observed in nature. In order to do this, a comparison with observations is obviously necessary. Observational studies of the distribution of delay times of SNe Ia can be used to constrain theoretical models for their progenitors and formation scenarios \citep[see e.g.][]{mannucci2006}. \citet{totani2008} selected 65 supernova candidates at redshifts from 0.4 to 1.2 (corresponding to a light travel time of 0.1 to 8.0 Gyr), of which they demonstrate that at least 80\% must be SNe Ia. These events were selected from old galaxies of the elliptical type, in which star formation has long ceased almost completely. Hence, the selected samples are equivalent to (passively evolving) starburst galaxies for the determination of the delay time of SNe Ia, and can be used for that purpose. The thus obtained observational delay time distribution (DTD) can then be directly compared to the theoretically predicted DTDs for starbursts. While \citet{totani2008} make the implicit assumption that the DTDs are independent of metallicity, the samples are expected to be contained within the range 1.0 to 2.5 times solar metallicity, so any metallicity effect should be limited and certainly less than an order of magnitude (as will be shown later). Also, it is assumed that the average age of local elliptical galaxies is 11 Gyr, and for those the authors adopt the SN Ia rate observed for such galaxies by \citet{mannucci2005}. A striking feature of the thus obtained observational DTD is that it follows a power law: it decreases inversely proportional to time (galaxy age).

The only caveat is that the observations are presented per unit delay time per century for a single starburst population whose total K-band luminosity $L_{K,0}$ at an age of 11 Gyr is $10^{10}$ L$_{\mathrm{K},\odot}$, while all population synthesis models give SN Ia rates in number of events per unit time for a single starburst with a given total initial mass. This implies that a conversion is required between SN Ia rate per K-band luminosity (SNuK) and SN Ia rate per mass (SNuM). \citet{totani2008} adopt a conversion factor obtained from a spectral energy distribution template for an exponentially decaying star formation rate with a timescale of 0.1 Gyr. This factor depends on the adopted metallicity $Z$ and initial mass function (IMF), and values are given for different choices. The cases assuming solar metallicity and the \citet{chabrier2003} IMF best match our use (see Sect. 2) of $Z=0.02$ and the \citet{kroupa1993} IMF. Following personal communication with Totani, we take a value based on a combination of these models:

\begin{equation}
\frac{M_*}{L_{K,0}} = 1.56\left[\frac{\mathrm{M}_{\odot}}{\mathrm{L}_{\mathrm{K},\odot}}\right].
\end{equation}

In this equation, $M_*$ is defined as the integrated star formation rate up to the age of the galaxy, which in the case of a single starburst is nothing else than the total initial mass. Since this conversion factor is taken to be constant for the entire redshift range, any change (or uncertainty) in it will only influence the absolute values of the SN Ia rate in SNuM, and not the morphological shape of the DTD in those units.

The question addressed in this work will thus pertain to the dominant formation scenario in the case of elliptical galaxies. As a novelty, we will specifically investigate the influence of the description of mass transfer efficiency during Roche lobe overflow (RLOF). In Sect. 2, we give a brief description of our population number synthesis code, and we elaborate on the new physical effects which have been incorporated as well as on the details of systems leading to a SN Ia. Sect. 3 contains our new results on the progenitors of SNe Ia, as well as a short elaboration on a special kind of merger. In Sect. 4, a comparison is made between the results of our code and those published by groups undertaking similar studies.

\section{Population number synthesis code}

For our study, we use the Brussels population number synthesis code developed over the years at the Astrophysical Institute of the Vrije Universiteit Brussel. A very extended review of the elements contained in this code is given by \citet{dedonder2004}. We repeat here only the most important points and the changes which have been implemented since then, reflecting recent developments in the description of certain physical processes.

Our code starts from an instantaneous starburst normalized to $10^6$ M$_{\odot}$ with $Z=0.02$ and a binary fraction of 100\%. By default the stars are drawn from a \citet{kroupa1993} IMF, flat mass ratio distribution and \citet{abt1983} initial orbital separation distribution. The implications of other choices will be discussed in Sect. 3. Supernova rates are given in the output as number of events per Gyr as a function of time elapsed since starburst.

\subsection{Treatment of mass transfer}

As stated above, a main new factor in this investigation is the influence of the description of the RLOF. Whether or not this process is conservative is indicated by the parameter $\beta$.

In the conservative case, $\beta = 1$, the total mass and angular momentum of the system are conserved during RLOF. Matter is transferred from the initially more massive star ($M_1$) to the initially less massive star ($M_2$), whereby the latter accretes all the matter lost by the donor. This process can continue after the donor star has become less massive than the accretor star, a phase known as mass ratio reversal. Angular momentum is exchanged between the two components, and with the orbit. The resulting orbital period is given by

\begin{equation}
\frac{P_\mathrm{f}}{P_\mathrm{i}} = \left(\frac{M_{1\mathrm{i}}M_{2\mathrm{i}}}{M_{1\mathrm{f}}M_{2\mathrm{f}}}\right)^3.
\end{equation}

The indices in this formula, i for initial and f for final, are with respect to the RLOF phase. This ratio is smaller or larger than unity depending on whether mass transfer continued until after mass ratio reversal.

When mass transfer is non-conservative, the following relation holds:

\begin{equation}
\beta = \left|\frac{\mathrm{d}M_{2}/\mathrm{d}t}{\mathrm{d}M_{1}/\mathrm{d}t}\right| < 1.
\end{equation}

$\beta$ is thus the fraction of material lost by the donor which is accepted by the accretor. The amount of mass lost from the system is readily found by integration over the time during which mass loss takes place. To determine the amount of angular momentum lost, and thus the final orbital period, however, requires knowledge of the way in which the non-accreted material leaves the system. The two most common approaches, which will be considered in this work, is that either the matter is lost with the specific angular momentum of the accretor, or with the orbital angular momentum of a corotating point at the second Lagrangian point ($\mathrm{L}_2$). The first assumption is justified if the mass loss can be treated as an enhanced stellar wind, or at least as a process which is symmetrical in the equatorial plane of the accretor. There is however no physical mechanism which is known to make this possible in normal intermediate mass binaries.

The formulas thus obtained for orbital period variation during non-conservative RLOF are the following \citep[see e.g.][]{dedonder2004}.

For $0 < \beta < 1$:

\begin{equation}
\frac{P_\mathrm{f}}{P_\mathrm{i}} = \left(\frac{M_{1\mathrm{f}}+M_{2\mathrm{f}}}{M_{1\mathrm{i}}+M_{2\mathrm{i}}}\right)\left(\frac{M_{1\mathrm{f}}}{M_{1\mathrm{i}}}\right)^{3\left[\sqrt{\eta}\left(1-\beta\right)-1\right]}\left(\frac{M_{2\mathrm{f}}}{M_{2\mathrm{i}}}\right)^{-3\left[\sqrt{\eta}\frac{1-\beta}{\beta}+1\right]}.
\end{equation}

For $\beta = 0$:

\begin{equation}
\frac{P_\mathrm{f}}{P_\mathrm{i}} = \left(\frac{M_{1\mathrm{f}}+M_{2\mathrm{f}}}{M_{1\mathrm{i}}+M_{2\mathrm{i}}}\right)\left(\frac{M_{1\mathrm{f}}}{M_{1\mathrm{i}}}\right)^{3\left(\sqrt{\eta}-1\right)}\mathrm{e}^{3\sqrt{\eta}\left(\frac{M_{1\mathrm{f}}-M_{1\mathrm{i}}}{M_{2\mathrm{i}}}\right)}.
\end{equation}

In the case where mass loss through the $\mathrm{L}_2$ point is assumed, the parameter $\eta$ is determined by the distance from this point to the center of mass, expressed in terms of the orbital separation. It was shown by \citet{dedonder2004} that a typical value is $\eta=2.3$. When the angular momentum loss is taken to be that specific to the accretor, $\eta$ can be much smaller than unity. The standard model will assume the former value of 2.3, the implications of other choices being briefly discussed.

When mass transfer becomes dynamically unstable, the binary will be enveloped in the extended atmosphere of the initially most massive component, resulting in what is known as a common envelope (CE) phase. The friction due to viscosity of the rotating pair within the envelope causes a significant loss of rotational energy, which leads to a strong reduction of the orbital period. This process is known as spiral-in. If the period is sufficiently reduced, both stars will merge. Otherwise, if there is sufficient binding energy released by the inspiraling pair, the envelope may eventually be ejected, again yielding a separated binary. Due to the short timescale of this process, it is assumed that during a CE phase the accretor does not gain any appreciable amount of mass, i.e. $\beta = 0$.

Different models exist for the description of energy conversion during CE. The $\alpha$-scenario by \citet{webbink1984} depends on the efficiency $\alpha$ with which orbital energy lost by the inspiraling stars can be converted into kinetic energy for the expulsion of the envelope. The change in orbital separation is given by

\begin{equation}
\frac{M_{1\mathrm{i}}\left(M_{1\mathrm{i}}-M_{1\mathrm{f}}\right)}{\lambda R_{\mathrm{Roche}}} = \alpha \left(\frac{M_{1\mathrm{f}}M_{2\mathrm{i}}}{2A_\mathrm{f}}-\frac{M_{1\mathrm{i}}M_{2\mathrm{i}}}{2A_\mathrm{i}}\right).
\end{equation}

In this equation, $R_{\mathrm{Roche}}$ is the Roche radius of $M_1$, while $\lambda$ is characterizing the density structure of this star's extended envelope and is dependent on whether this envelope is radiative or convective. By definition, $\alpha$ must lie between 0 and 1, while it is also conceivable that the value is dependent on the nature of the object which is performing the spiral-in. The resulting orbital period variation is obtained using Kepler's third law. When we apply this scenario in our code, the standard choice will be that $\alpha\lambda=1$, where the Roche lobe is calculated using the fit by \citet{eggleton1983}. The implications of different choices will be described below.

An alternative description of the CE evolution phase is given by \citet{nelemans2005}. The change in angular momentum is described by

\begin{equation}
\frac{\Delta J}{J} = \gamma\frac{\Delta M_{\mathrm{total}}}{M_{\mathrm{total}}} = \gamma\frac{M_{1\mathrm{i}}-M_{1\mathrm{f}}}{M_{1\mathrm{i}}+M_{2\mathrm{i}}}
\end{equation}

where $\gamma$ is a parameter with a typical value around 1.5. Using this formalism, the change in orbital separation is as follows:

\begin{equation}
\frac{A_\mathrm{f}}{A_\mathrm{i}} = \left(\frac{M_{1\mathrm{i}}}{M_{1\mathrm{f}}}\right)^2\left(\frac{M_{1\mathrm{f}}+M_{2\mathrm{i}}}{M_{1\mathrm{i}}+M_{2\mathrm{i}}}\right)\left(1 - \gamma \frac{M_{1\mathrm{i}}-M_{1\mathrm{f}}}{M_{1\mathrm{i}}+M_{2\mathrm{i}}}\right)^2.
\end{equation}

This equation can again be readily converted into one for the evolution of the orbital period using Kepler's third law. The authors argue that this formalism is to be preferred in systems which will lead to the formation of a double WD or WD+MS/RG binary. In brief, this $\gamma$-scenario starts from a conservation of energy and a balance of angular momentum, whereas the $\alpha$-formalism assumes a conservation of angular momentum to arrive at a balance of energy.

\subsection{Single degenerate scenario progenitors}

\citet{hachisu1996} introduce a SD SN Ia progenitor channel consisting of a WD accreting hydrogen rich material from a low mass RG. High mass accretion rates lead to the development of a strong wind, which stabilizes the mass transfer rate. The WD thus steadily accretes matter, which is first burned into helium, without the formation of a CE, allowing it to reach the Chandrasekhar mass. \citet{li1997} extended this scenario to MS companions. \citet{hachisu1999b} give a scenario which takes into account both possibilities for $Z=0.02$. For different WD masses, regions in the orbital period - companion mass parameter space are presented which will eventually result in a SD SN Ia. There are two such `islands' in the ($P$, $M_{\mathrm{companion}}$)-plane: one for (late) MS companions, ranging from about 2 to 3 M$_{\odot}$ companions with orbital periods of about 0.4 to 6 days, and a second one for RG companions, ranging from about 1 to 3 M$_{\odot}$ and 30 to 300 days. The first channel is known as the WD+MS channel, the second as the WD+RG channel. If at any time between the formation of the first and second WD in our potential progenitors, the system enters one of these contours, it is assumed that it will eventually result in a SN Ia through the SD scenario. The explosion is taken to occur at the time that the mass transfer toward the WD starts, as the extra time which is required in reality is negligible compared to the nuclear lifetime of both stars. \citet{hachisu2008} introduce the influence of the mass stripping effect on the WD+MS channel. This effect, resulting from the interaction of the WD wind with the surface of the companion, attenuates the mass transfer rate from the MS star onto the WD. MS companions can thus be more massive than previously expected, and still avoid a CE phase, allowing them to remain a potential SD SN Ia progenitor. The fraction of mass stripped away by this wind is

\begin{equation}
\frac{\mathrm{d}M_{\mathrm{strip}}}{\mathrm{d}t}=c_1\frac{\mathrm{d}M_{\mathrm{wind}}}{\mathrm{d}t}.
\end{equation}

$c_1$ is a parameter depending on, among others, the wind velocity, the efficiency of conversion from kinetic energy to thermal energy, and the geometrical factor of the surface hit by the wind. The value of this parameter is quite uncertain, and best fit models based on observations yield a possible range from 1 to 10. A brief discussion on the implications of varying this parameter will be given, while a moderate value of $c_1 = 3$ will be adopted for the standard parameter set. In that case, the WD+MS channel contour typically extends up to a companion mass of 6 M$_{\odot}$, instead of 2 M$_{\odot}$ without the effect.

\citet{kobayashi1998} investigated the influence of metallicity on the SD scenario. They find that if the accretor's iron abundance is too low, its wind is too weak to allow stable mass accumulation up to the Chandrasekhar mass, and hence no SN Ia can occur. The area in the parameter space which leads to SD events thus decreases with metallicity. We use their progenitor regions in the ($P$, $M_{\mathrm{companion}}$)-plane for $Z=0.004$ together with those of \citet{hachisu1999b,hachisu2008} to interpolate for any metallicity.

\subsection{Double degenerate scenario progenitors}

In the DD scenario, it is assumed that every WD merger exceeding 1.4 M$_{\odot}$ results in a SN Ia. In our code, which follows the evolution of the progenitor in detail throughout its existence, without the use of an analytical formalism, there are two typical formation channels which lead to a SN Ia. We have termed them the RLOF channel and the CE channel. A description as well as a typical example of both is given below, since experience has shown that there exists a great diversity in evolutions obtained with different codes (see also Sect. 4). The numbers in this subsection are given for a somewhat stricter subset of the progenitors of DD SNe Ia, i.e. those in which both merging WDs are of the C-O type. If this requirement is waived, the parameter space of progenitors is slightly extended. The implications of this limitation on the eventual DTDs will be discussed in the Results section.

\subsubsection{The Roche lobe overflow channel}

In this channel, the progenitor system at birth consists of a primary typically between 3.7 and 7.1 M$_{\odot}$, a mass ratio between 0.35 and 1.0 (high mass ratios being strongly favored in case of all but the highest primary masses), and an orbital period of up to 100 days. After the most massive component evolves off the MS and traverses the Hertzsprung gap, a first phase of mass transfer will take place. The envelope of the donor is radiative at that time, resulting in a dynamically stable mass transfer on a nuclear or thermal timescale, which is a canonical RLOF. The standard model is to treat this RLOF conservatively ($\beta = 1$), i.e. all matter lost by the donor is accreted by the other star. The mass transfer will continue until well after mass ratio reversal, resulting in a system with a more extreme mass ratio (where the donor is now the least massive star) and an increased orbital period, typically a few hundred days. The evolution of the accretor under the effect of mass accretion (i.e. rejuvenation) has been explicitly calculated by \citet{vanbeveren1998}. This has been done with a standard accretion model \citep[see][]{neo1977} if matter directly impacts the surface of the star, and under the assumption of accretion induced full mixing \cite[see][]{vanbeveren1994} if a Keplerian accretion disk is formed. During the collapse of the former donor to a WD, a fraction of its mass will be expelled, resulting in a somewhat larger period still. After the initially least massive star eventually also leaves the MS (a timescale which is critically affected by the amount of mass it received during the first mass transfer phase), it will also increase in radius until it exceeds its Roche lobe. Thus, a second mass transfer phase, toward the WD, will be initiated. Due to the extreme mass ratio of the system and the fact that the accretor is a WD, this mass transfer will become dynamically unstable, resulting in a CE phase leading to a spiral-in. Eventually, the envelope will be ejected and a double WD binary will emerge. This system consists of two WDs with a mass around one solar mass each and an orbital period which has decreased dramatically to a few hours during the spiral-in. Constant emission of GWR will cause the two components to slowly spiral in over the course of a few billion years. Eventually, they will merge, and since they together exceed the Chandrasekhar mass, will under our assumptions result in a SN Ia. The typical timescale of this event is the sum of the formation timescale of the double WD binary and the GWR spiral-in. This is typically on the order of a few Gyr, but can be lower for systems which had very short initial periods, resulting in a delay time varying between 0.22 Gyr up to the Hubble time and beyond.

If the same initial system is considered, but the first mass transfer phase is taken to be totally non-conservative ($\beta=0$) the system will merge before it can result in a double WD binary, and thus there will not be a SN Ia explosion.

\paragraph{Example}

We start from a system on the ZAMS consisting of 4.0+3.6 M$_{\odot}$ stars with an initial orbital period of 5.0 days. After the first mass transfer phase at 0.20 Gyr, which is a canonical RLOF and is considered to be conservative, a 0.65+7.0 M$_{\odot}$ system emerges with a period of 160 days. Just prior to the originally least massive star filling its Roche lobe, the other star has become a WD with a mass of 0.57 M$_{\odot}$ and the period has increased to 170 days. This second mass transfer phase at 0.26 Gyr results in a CE evolution (described with the $\alpha$-formalism), eventually yielding a system of 0.57 + 0.88 M$_{\odot}$ WDs with a period of 0.19 days. This requires 3.1 Gyr of GWR in order to merge, resulting in a DD SN Ia after 3.4 Gyr in total. The evolution of this system is represented in the left panel of Figure 1. Using the same initial conditions but assuming a totally non-conservative RLOF as first mass transfer phase, the system already merges during that phase, and thus does not result in a SN Ia.

\begin{figure*}
\centering
   \includegraphics[width=12cm]{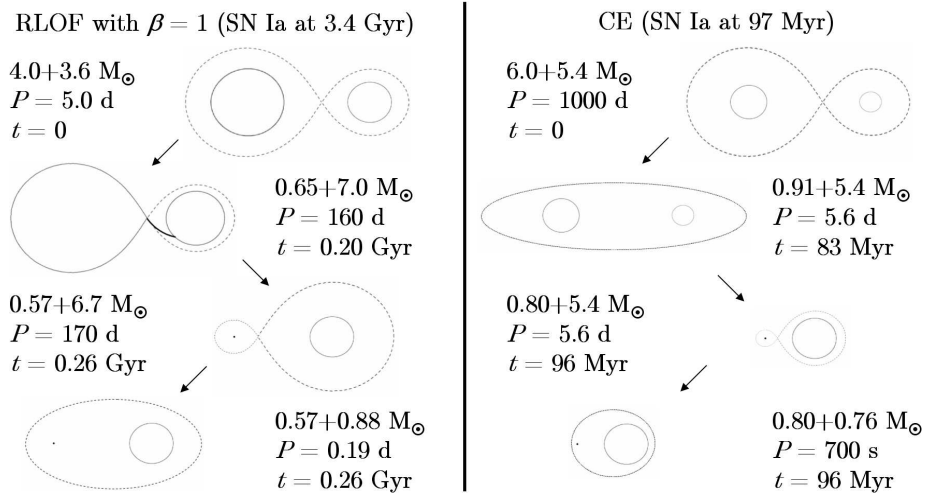}
     \caption{A graphical representation (not to scale) of the evolution of a system leading to a DD SN Ia through the RLOF channel (left panel) and the CE channel (right panel).}
\end{figure*}

\subsubsection{Common envelope channel}

Alternatively, one can start with a system which is quite similar in initial masses, i.e. a primary mass between 5.3 and 7.9 M$_{\odot}$ and a mass ratio between 0.84 and 1.0, but with a much larger initial orbital period, starting at 200 days. The large orbital separation will ensure that the most massive star does not exceed its Roche lobe until after it has developed a deep convective envelope. The result is that this mass transfer phase will be dynamically unstable, and will already here result in a CE spiral-in. This will decrease the orbital period by about two orders of magnitude. After the envelope is ejected, a binary with a less extreme mass ratio than in the RLOF scenario (the accretor has not gained any material due to the fast timescale of the mass transfer) and an orbital period of only a few days is obtained. When the originally most massive star has become a WD and the other has started evolving off the MS, a second mass transfer phase will be initiated. Just as before, this will result in a CE phase and spiral-in. Systems which do not merge during this phase will eventually emerge as a double WD binary of the order of one solar mass each, with an orbital period of only a few tens of seconds. GWR has to work only a few tens of thousands of years in order to let such a system merge, which means that the resulting SN Ia is not delayed much more than the formation timescale of the double WD binary. This typically amounts to somewhat more than a hundred million years, but can vary from 55 to 180 Myr depending on the initial conditions.

\paragraph{Example}

An initial system with 6.0+5.4 M$_{\odot}$ stars and an initial orbital period of 1000 days encounters a CE phase during the first mass transfer at 83 Myr, emerging as a 0.91+5.4 system with a period of 5.6 days. The second mass transfer phase at 96 Myr, toward a 0.80 M$_{\odot}$ WD, also results in spiral-in. The eventual result is a 0.80+0.76 M$_{\odot}$ WD binary with a period of 700 s, which needs 650 kyr of GWR and results in a DD SN Ia 97 Myr after star formation. This evolution is rendered in the right panel of Figure 1.

\subsection{White dwarf plus main sequence or red giant mergers}

If at any time during the evolution, after the originally most massive star has become a WD but while the other is still non-degenerate, the components merge, the result is a priori unknown. \citet{sparks1974} suggested that this could lead to a supernova explosion. Our code calculates the number of such mergers for which the combined mass of the WD and the companion's core exceeds the Chandrasekhar mass. As the MS or RG star still has hydrogen in its outer layers, we will tentatively classify such events as type II supernovae. It is however not yet known whether this is the case in nature.

\section{Results and discussion}

\subsection{Single versus double degenerate scenario}

The DTDs which were obtained with the described code are rendered in Figure 2. This shows the DTDs using the $\alpha$-scenario for CE evolution ($\alpha=1$) and with $\beta = 1$, for both the SD and DD model. Also represented are the observational data points, with error bars, from \citet{totani2008}. The observations have been converted from SN Ia rate per K-band luminosity into SN Ia rate per mass using Eq. (1). This includes the one by \citet{mannucci2005} for local ellipticals, which are assumed to have an age of 11 Gyr.

\begin{figure*}
\centering
   \includegraphics[width=12cm]{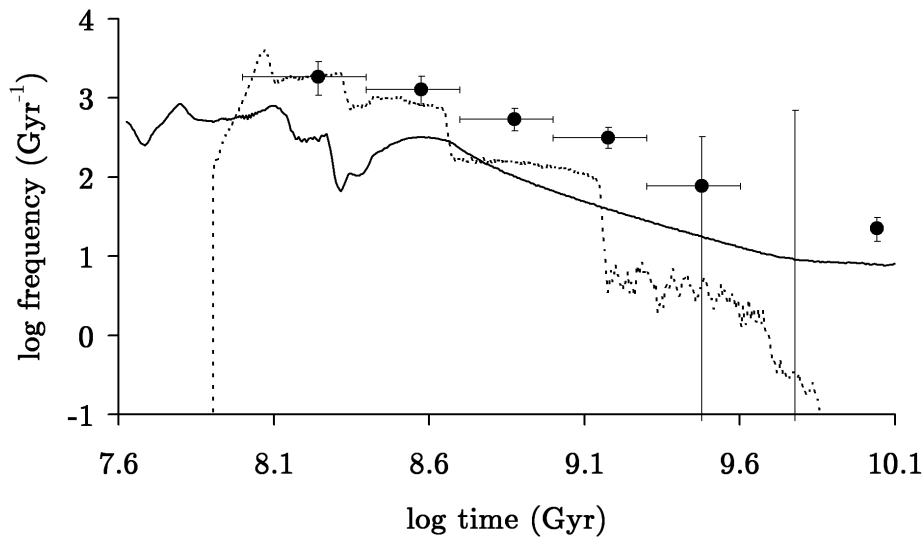}
     \caption{DTDs of a $10^6$ M$_{\odot}$ starburst obtained for the DD (solid) and SD (dotted) scenario with $\beta = 1$ and using the $\alpha$-formalism for CE-evolution. Observational data points of \citet{totani2008} and \citet{mannucci2005}.}
\end{figure*}

The most important observations are the following: \\
$\bullet$ As is clear from the figure, during most of the time after star formation, the DD DTD is dominant over the SD one, at later times by an order of magnitude or more. For much of the first 1.4 Gyr, the SD DTD shows a spike which causes it to be dominant during this time, also by up to about an order of magnitude. After these early events however, the SD DTD decreases rapidly, then levels off and after about 8 Gyr disappears completely. This cutoff time is determined by the nuclear lifetime of the least massive companion stars that can lead to a SD SN Ia. The SD DTD alone is thus clearly incompatible with the morphological shape dictated by the observational data points. \\
$\bullet$ The DD DTD follows a decay inversely proportional to time, the shape of which is in fair agreement with that of the data points. A DTD made up of the sum of events caused through both the SD and DD scenario matches most observational points even better in morphological shape, apart from those at very early times. It is indeed possible that both models should be considered together, since they originate from two distinct groups of progenitor systems, whose initial parameters do not overlap. \\
$\bullet$ In terms of absolute number of events, neither the SD scenario, DD scenario, or a combination of both is able to produce enough SNe Ia to match the observed values, especially at later times. At the 11 Gyr point the discrepancy amounts to a factor of three, a number which would even increase if the initial binary fraction is taken smaller than 100\%. Similar conclusions have already been reached by other groups performing theoretical DTD studies (see Sect. 4), e.g. \citet{ruiter2009} who found a rate almost ten times smaller than the SNuM given by \citet{mannucci2005}. A possible explanation is that the constant conversion of the observational results from SNuK into SNuM may introduce large systematic errors, which is visible from the uncertainties on the conversion factor as well as the difference in its values given by \citet{mannucci2005} and \citet{totani2008}. As a result, the remainder of this section will mainly concentrate on a comparison between the morphological shapes of theoretical and observational DTDs, not on their absolute values. \\
$\bullet$ We find that nearly all SD events are created through the WD+MS channel described above. \\
$\bullet$ A great majority of DD SNe Ia (about 80\%) are formed through a RLOF phase followed by a CE phase, as opposed to two successive CE phases. This will be further discussed in the next subsection.

\subsection{Mass transfer efficiency}

In Figure 3, the influence of mass transfer efficiency is investigated. It shows the DD DTD for different values of accreted mass fraction $\beta$. The figure shows that in the case of totally non-conservative RLOF ($\beta = 0$), the DTD drops dramatically after 0.2 Gyr, leaving virtually no events beyond 0.5 Gyr. This directly implies that in the case of a non-conservative first mass transfer phase, there will be no DD SNe Ia with a sizable delay time. Since non-conservative RLOF excludes systems going through such a phase as SN Ia progenitor, the events which do remain in this case (all with short delay time) must be produced through the double CE channel. On the other hand, the events which are absent in the case of $\beta = 0$, but account for about 80\% of all DD SNe Ia in the case of conservative RLOF, are those created through a RLOF phase followed by a CE evolution. These results can be understood in terms of the two typical evolution channels leading to a DD SN Ia discussed before. It was shown that the RLOF channel (one RLOF followed by one CE phase) produces events with a delay time of multiple Gyr, but is unable to produce any SN Ia if RLOF is non-conservative. The CE channel (two successive CE phases) on the other hand is independent of $\beta$, but can only produce events during the first few hundreds of millions of years after starburst.

\begin{figure*}
\centering
   \includegraphics[width=12cm]{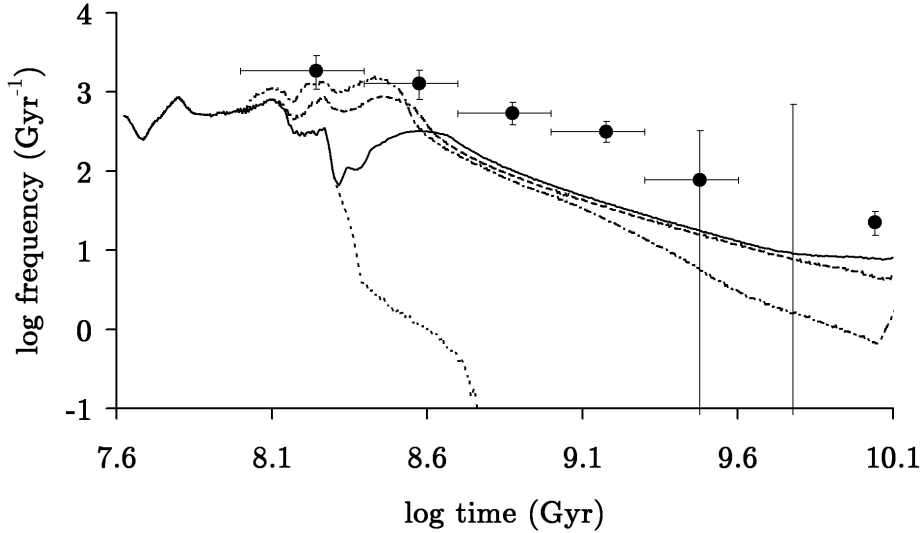}
     \caption{DTDs of a $10^6$ M$_{\odot}$ starburst obtained for the DD scenario with $\beta = 1$ (solid), $\beta = 0.9$ (dashed), $\beta = 0.8$ (dashed dotted) and $\beta = 0$ (dotted) using the $\alpha$-formalism for CE-evolution. Observational data points of \citet{totani2008} and \citet{mannucci2005}.}
\end{figure*}

In order for the model to be consistent in DTD shape with observations, a nearly conservative mass transfer is necessary, which means that $\beta$ needs to be close to one overall. Further investigation has pointed out that, on average, $\beta \ge 0.9$ is required for the theoretical model to remain marginally consistent in DTD shape with the observations under the current assumptions. This is also visible from Figure 3. An important consequence is that taking the formation timescale of a double WD binary equal to the MS lifetime of the least massive component is unjustified. This practice, commonly adopted in studies which make use of analytical formalisms instead of detailed evolution, is namely based on the unfounded assumption that the accretor is not affected by the mass transfer process, either because it is totally non-conservative or a CE evolution (which is by definition taken to be too fast for the accretor to gain any mass).

The description of angular momentum loss does not affect this quasi-conservative model much, since there is only very little mass which leaves the system. However, when calculations are made with lower values of $\beta$, the DTD is critically affected by the choice for this description. Mass loss through the corotating ring causes a DTD which descends much steeper than with the assumption of specific accretor angular momentum loss. This implies that the latter description allows a broader range of $\beta$ to be compatible in DTD shape with the observations (see also Sect. 4).

The SD DTD has also been calculated with $\beta = 0$, but in that case there is little difference with the result shown in Figure 2: the DTD drops only very slightly, since the value of $\beta$ is of only small influence on the number of progenitor systems which will at some point lie within the zones identified by \citet{hachisu2008}. This is because some 90\% of these systems have gone through a CE phase during the first mass transfer, and are thus not affected by the degree of conservatism of stable RLOF.

All these results were obtained under the assumption that all WD mergers exceeding 1.4 M$_{\odot}$ result in a SN Ia. If this assumption is changed into including only the merger of C-O WDs satisfying the same criterion, the absolute number of DD SNe Ia decreases by about 40\%, and the separation in time between RLOF and CE events thus becomes more distinct. However, none of the above conclusions are altered in any way, since the overall shape of the DTDs does not change.

\subsection{Influence of other parameters}

The influence of the $c_1$ parameter in the mass stripping effect is investigated next. So far, $c_1=3$ represented the standard model favored by \citet{hachisu2008}. When $c_1$ is taken equal to 0, the effect is essentially turned off, and the earliest and highest spike in the SD DTD disappears. This makes the absolute number of SD SNe Ia during early times ($<$ 0.5 Gyr) no longer exceeding those of the DD channel, effectively making the latter dominant over the entire time range. Taking $c_1>3$, up to 10 according to the empirically obtained values, further enhances the peak which is present for $c_1=3$, but not by a significant amount.

Results discussed so far have been obtained with the $\alpha$-scenario of \citet{webbink1984} for CE evolution. When the $\gamma$-scenario by \citet{nelemans2005} is used ($\gamma=1.5$), the results obtained for $\beta = 1$ are shown in Figure 4. It is obvious that the SD DTD still has a morphological shape which is incompatible with the observational data points: it drops away much too soon and too fast. Whereas the DD DTD lies almost an order of magnitude lower in absolute rate than when using the $\alpha$-scenario, this result can by no means be rejected based on a comparison with the morphological shape of the observational DTD. An interesting point is that when $\beta$ is taken smaller than one, the number of SNe Ia in both channels decreases (as with the $\alpha$-scenario), albeit the DD DTD does not disappear as dramatically as before. For $\beta = 0$ it drops by only about 10\%, rendering a broad range of $\beta$ compatible in DTD shape with observations. The reason is that with the $\gamma$-scenario, there are much more systems which are formed through a double CE phase, and are thus not affected by the value of $\beta$. It is very important however to remark that the validity of the $\gamma$-scenario is questionable due to the fact that almost no events predicted by it result from the merger of two C-O WDs. If progenitors are restricted to such systems, adoption of the $\gamma$-scenario will thus hardly yield any DD SNe Ia.

\begin{figure*}
\centering
   \includegraphics[width=12cm]{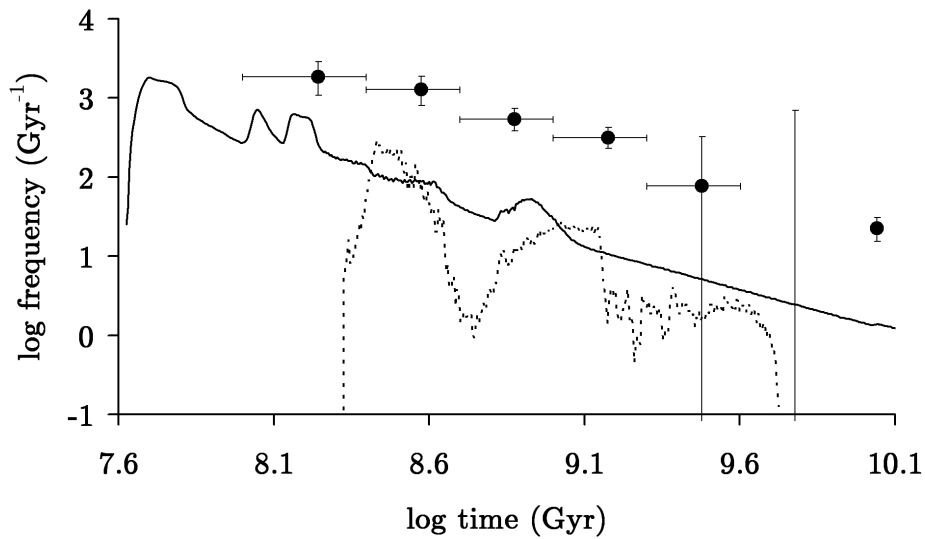}
     \caption{DTDs of a $10^6$ M$_{\odot}$ starburst obtained for the DD (solid) and SD (dotted) scenario with $\beta = 1$ and using the $\gamma$-formalism for CE-evolution. Observational data points of \citet{totani2008} and \citet{mannucci2005}.}
\end{figure*}

When the $\alpha$-scenario is considered again, but this time with $\alpha=0.5$ instead of $\alpha=1$, the shape of both DTDs and the absolute number of events change, but not in a way to alter conclusions. The DD DTD comes to lie higher (+60\% in absolute number of SNe Ia), but is still consistent in shape with observations, possibly in combination with the SD DTD. The latter one drops slightly compared with $\alpha=1$ (-30\%), but does not change in overall shape. The dependence of the DTDs on the value of $\beta$ is similar to that in the case of $\alpha=1$, but with an even earlier cutoff for the DD DTD in the case of $\beta=0$.

Also the influence of initial mass ratio distribution is investigated. Using a different distribution than the flat one used so far \citep[e.g. the ones by][]{garmany1980,hogeveen1992} results in a change of the absolute number of SNe Ia in both scenarios, and can also slightly affect the morphological shape of the theoretical DTDs. However, the change is too insignificant to alter the conclusions above. The same is true for the choice of IMF: until now, we have used the one by \citet{kroupa1993}, and taking another one \citep[e.g. following][]{salpeter1955} does have an effect on the absolute number of events, but not on the shape of the DTDs or the conclusions made so far.

A final parameter which is scrutinized is the metallicity of the galaxy. So far, calculations have been made with a solar $Z=0.02$, since this is the order of magnitude at which the available observations have been made. If we take a Magellanic Cloud metallicity of $Z=0.002$, the DD DTD is not much affected in shape and increases by about 50\% in number, whereas the SD DTD significantly drops over its entire range, leaving only 3\% of the absolute number of SD SNe Ia. This is a direct result of the fact that the areas in the ($P$, $M_{\mathrm{companion}}$)-plane of systems which will result in a SD SN Ia decrease dramatically with metallicity. An interesting consequence is that very old galaxies with $Z$ close to zero are unable to produce a sizable number of SNe Ia through the SD scenario.

In this respect, it should be remarked that our previous conclusions are not compromised by the combination of this metallicity effect and the fact that the observational data points were taken from samples with a metallicity of up to 2.5 times the solar value. The reason is that for $Z = 0.05$ the theoretical SD DTD stays within an order of magnitude of the $Z=0.02$ one, and its shape also strongly deviates from that of the observational data points.

\subsection{White dwarf plus main sequence or red giant mergers}

Figure 5 shows the DTD for the merger of a WD and a MS star or RG. This demonstrates that for both extreme values of $\beta$, these (potential) special type II supernovae start very soon after starburst (about 20 Myr) and extend over a few hundred Myr. They originate from systems with a wide range of initial primary mass (above 4.5 M$_{\odot}$), mass ratio (between 0.2 and 1.0) and orbital period. In the case of conservative RLOF ($\beta=1$), about one quarter of the events are formed through two successive CE phases. These are the only ones which remain for $\beta=0$.

\begin{figure*}
\centering
   \includegraphics[width=12cm]{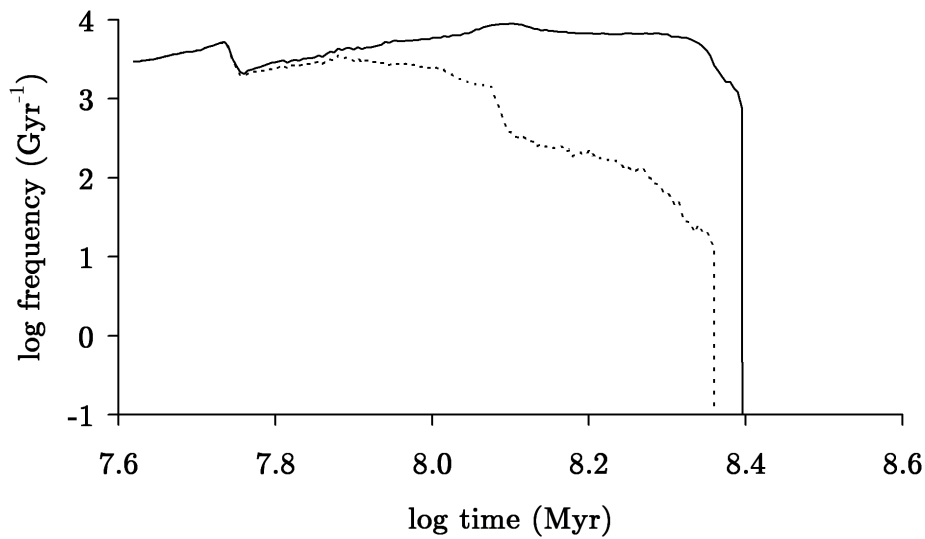}
     \caption{DTDs of a $10^6$ M$_{\odot}$ starburst obtained for WD+MS/RG mergers with $\beta = 1$ (solid) and $\beta = 0$ (dotted) using the $\alpha$-formalism for CE-evolution.}
\end{figure*}

\section{Comparison with previous studies}

The results of our population number synthesis can, after a simple conversion of units, be directly compared to results obtained by other groups in similar theoretical population studies. As sketched above, population synthesis models are very sensitive to a large number of input parameters, such as mass transfer efficiency, CE description, angular momentum loss mechanism, initial mass, mass ratio and orbital period distribution, and others. In order to carry out the following comparisons, we have equipped our code with the assumptions, parameters and initial conditions of the other groups, as far as the code technically allows. The descriptions of the other codes are given in the next subsections, while the comparison with ours is made in Sect. 4.5.

\subsection{Yungelson \& Livio}

The first study with which a comparison is made is that of \citet{yungelson2000}. The authors give, among others, a study of the DD scenario. Most assumptions were obtained through personal communication with Yungelson. The binary fraction is set to 100\%, while an IMF with exponent -2.5 is assumed. For the initial mass ratio distribution a broken power law is used, which is flat for close binaries. The distribution of initial orbital separations is taken to be logarithmically flat. A CE formalism using the $\alpha$ parameter with $\alpha=1$ is used, however with an important change in the description:

\begin{equation}
\frac{\left(M_{1\mathrm{i}}+M_{2\mathrm{i}}\right)\left(M_{1\mathrm{i}}-M_{1\mathrm{f}}\right)}{A_\mathrm{i}} = \alpha\left(\frac{M_{1\mathrm{f}}M_{2\mathrm{i}}}{A_\mathrm{f}}-\frac{M_{1\mathrm{i}}M_{2\mathrm{i}}}{A_\mathrm{i}}\right).
\end{equation}

This equation yields eventual orbital periods which are typically a factor of four larger than those obtained with Eq. (6). The eventually given DTD is for an instantaneous starburst, normalized to a formation of 4.7 M$_{\odot}$ of stars per year.

\subsection{Han \& Podsiadlowski}

\citet{han2004} only consider the SD channel for the formation of SNe Ia. Within this scenario, only progenitors which consist of a WD accreting from an unevolved or slightly evolved non-degenerate star are considered, i.e. the WD+MS channel in our terminology. They employ the Eggleton stellar evolution code for a $Z = 0.02$ population. A critical mass transfer rate toward the WD is given, above which no accretion takes place and mass transfer becomes non-conservative:

\begin{equation}
\left(\frac{\mathrm{d}M}{\mathrm{d}t}\right)_{\mathrm{crit}}=5.3\cdot10^{-7}\frac{1.7 - X}{X}\left(M_{\mathrm{WD}}-0.4\right).
\end{equation}

In this equation, $X$ is the hydrogen mass fraction and masses and rates are expressed in M$_{\odot}$ and M$_{\odot}$/yr respectively. Mass loss from the system is taken to carry away the specific angular momentum of the WD. A \citet{miller1979} IMF, constant initial mass ratio distribution and logarithmically flat initial orbital separation distribution are adopted, while limits on mass, mass ratio and orbital period distributions are given. About 50\% of systems are binaries with an orbital period below 100 years. For CE evolution, the $\alpha$-scenario is used with different values for $\alpha$. A DTD is given for an instantaneous starburst of $10^{11}$ M$_{\odot}$. No mass stripping is taken into account.

\subsection{Hachisu et al.}
In the same paper which describes the SD channel, \citet{hachisu2008} use this scenario to carry out a population study themselves. They only consider the SD scenario, through both formation channels, for a starburst with an exponent -2.5 IMF, flat mass ratio distribution and logarithmically flat orbital period distribution. Unfortunately, only relative rates are given and the time binning is quite crude. As a result, only the morphological shape of this DTD can be compared to ours.

Therefore, we also make a comparison with the results of \citet{hachisu1999a} as presented by \citet{han2004}. These do however not include the WD+RG channel (through which we do not obtain a significant number of SNe Ia anyway), nor the mass stripping effect for the WD+MS channel. In this study the mass transfer rate toward the WD is taken to be

\begin{equation}
\left|\frac{\mathrm{d}M_2}{\mathrm{d}t}\right|=\frac{M_2}{\tau_\mathrm{KH}}\mathrm{Max}\left(\frac{\zeta_\mathrm{RL}-\zeta_\mathrm{MS}}{\zeta_\mathrm{MS}},0\right).
\end{equation}

In this equation, $\tau_\mathrm{KH}$ is the Kelvin-Helmholtz timescale, $\zeta_\mathrm{RL}$ the mass-radius exponent of the inner critical Roche lobe, and $\zeta_\mathrm{MS}$ that of the MS star. The resulting DTD is given by \citet{han2004}, using the same initial conditions as for their own model.

\subsection{Ruiter et al.}

The final study with which our results will be compared is that of \citet{ruiter2009}. This work investigated both the SD and DD scenario using the evolution code of \citet{belczynski2008}. This is done for a metallicity of 0.02, a binary fraction of 50\%, a \citet{kroupa1993} broken power law for the IMF with given mass range, a flat mass ratio distribution and an \citet{abt1983} initial orbital separation distribution. For non-degenerate accretors, half of the mass that is transferred from the donor is accreted, while the other half is lost from the system with the specific angular momentum of the accretor. In the case of WD accretors, mass transfer is by default conservative, however the accretion rate is Eddington-limited. The $\alpha$-scenario for CE evolution is used, with discrete values for $\alpha\lambda$. The DTD is given both for an instantaneous burst of star formation and a constant star formation rate during 10 Gyr.

\subsection{Discussion}

Adopting the elements of the respective codes described above, we carried out DTD calculations with the Brussels code. We find a good agreement with all of them, both in DTD shape and absolute number of SNe Ia. A comparison with the output obtained using our own assumptions (with $\beta=1$ and the $\alpha$-formalism) is shown in Figure 6 for the DD scenario and in Figure 7 for the SD scenario. There are a few differences, which are discussed below: \\
$\bullet$ Our SD DTD shows a very high but short spike at very short delay times, which in shape matches that of the WD+MS events by \citet{hachisu2008}. We do however not reproduce the significant number of events they find through the WD+RG channel, since we do not have a non-negligible number of WD systems traversing this progenitor region in the parameter space. \\
$\bullet$ The aforementioned early spike is not present in the studies of \citet{han2004} and \citet{hachisu1999a}. As explained before, this is the result of the inclusion of mass stripping in our WD+MS channel, which was not done in those two studies. When we take $c_1 = 0$, i.e. turn the effect off as well, the spike disappears and the early SD DTD is in good agreement with theirs. We do however retain a low background of SD SNe Ia beyond 1.4 Gyr which is not present in the comparison ones. \\
$\bullet$ This low continuing SD rate is however confirmed by \citet{ruiter2009}, but these are the only events they find: the early maximum is not present is their study. A possible explanation for this is that \citet{ruiter2009} report that most of their SD events are created through the WD+RG channel (including subgiants) as opposed to the WD+MS channel, whereas in our case the opposite is true. It thus seems that the early maximum which is for some reason absent in their study corresponds to these events. \\
$\bullet$ As far as the DD DTD is concerned, we find a very good agreement with both \citet{yungelson2000} and \citet{ruiter2009}. For the latter, this is however only true for $\beta=0.5$ (which corresponds to their mass loss mechanism for RLOF toward a non-degenerate accretor) if we adopt a very small value for $\eta$ ($\le0.25$, simulating their assumption of specific accretor angular momentum loss).

\begin{figure*}
\centering
   \includegraphics[width=12cm]{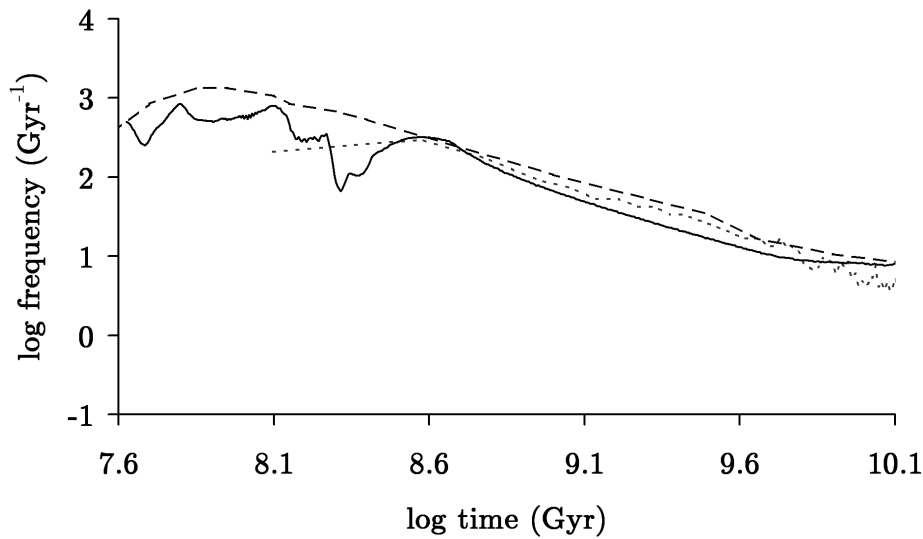}
     \caption{DTD of a $10^6$ M$_{\odot}$ starburst obtained for the DD scenario with $\beta = 1$ and using the $\alpha$-formalism for CE-evolution, compared with various other theoretical studies (see text). The solid line is the DTD from the present study, the dotted one is from \citet{ruiter2009} and the dashed one from \citet{yungelson2000}.}
\end{figure*}

\begin{figure*}
\centering
   \includegraphics[width=12cm]{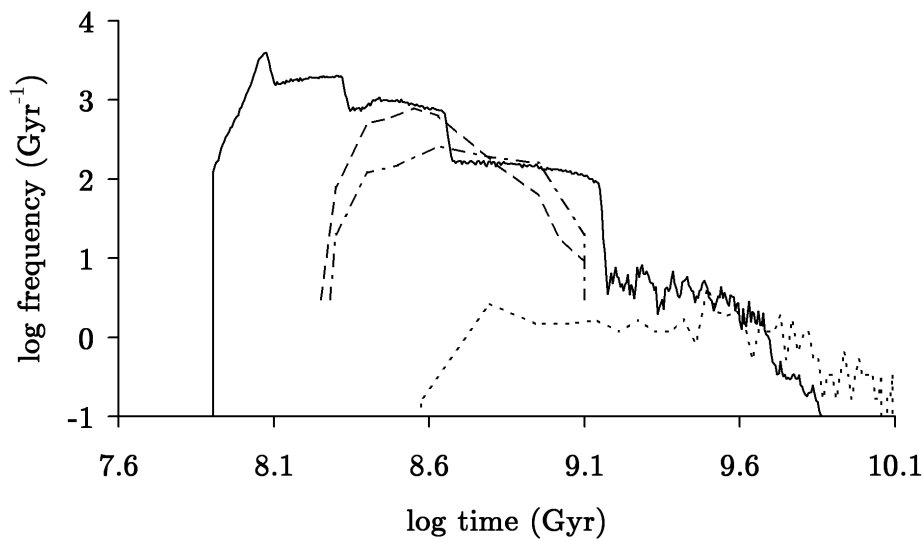}
     \caption{DTD of a $10^6$ M$_{\odot}$ starburst obtained for the SD scenario with $\beta = 1$ and using the $\alpha$-formalism for CE-evolution, compared with various other theoretical studies (see text). The solid line is the DTD from the present study, the dotted one is from \citet{ruiter2009}, the dashed one from \citet{hachisu1999a} and the dashed dotted one from \citet{han2004}.}
\end{figure*}

\section{Further considerations and outlook}
In Sect. 3, a discrepancy in absolute number between theoretical and observed SN Ia rates was discussed. A possible solution to this is stellar rotation: there seem to be indications \citep[see e.g.][]{habets1989} that it is common for components in binaries to rotate faster at birth than average single stars, which makes it possible that a significant fraction of primaries rotate faster than synchronous. If so, their MS convective core masses will be larger \citep[see e.g.][]{maeder2000,decressin2009}. Hence, the same will be true for their remnant masses after RLOF and the number of double WD binaries which in total mass exceed the Chandrasekhar mass will increase, leading to more SNe Ia through the DD scenario. We find that a convective core mass increase of only 10\% is enough for the theoretical rate at 11 Gyr to match the observed one. The thus obtained DTD, for both the SD and DD scenario combined, is rendered in Figure 8. This figure also includes the very recent observational DTD as reported by \citet{maoz2010}.

\begin{figure*}
\centering
   \includegraphics[width=12cm]{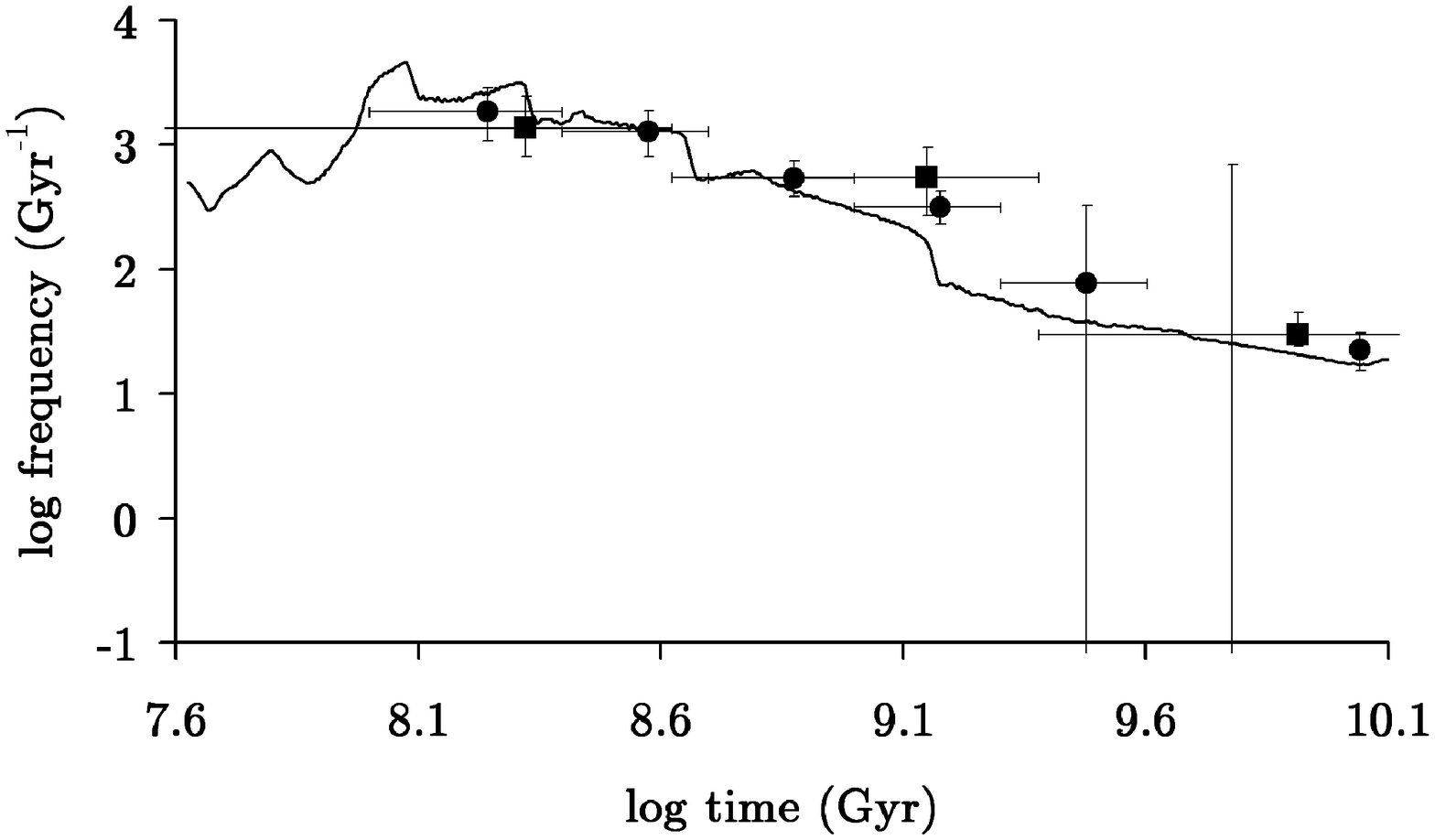}
     \caption{DTD of a $10^6$ M$_{\odot}$ starburst obtained for a combination of the SD and DD scenario with $\beta = 1$, using the $\alpha$-formalism for CE-evolution and with a 10\% increase in MS convective core masses. Observational data points of \citet{totani2008} and \citet{mannucci2005} (circles), as well as \citet{maoz2010} (squares).}
\end{figure*}

Also recently, \citet{wang2010} proposed a SD mechanism in which instabilities in an accretion disk formed around the WD can lead to SNe Ia with very long delay times, of up to the Hubble time. However, the absolute number of events thus obtained at long delay times (beyond 3 Gyr) lies about an order of magnitude or more below that obtained through the DD scenario in the current paper, and hence even further below the observations. As a result such mechanism, while beyond the scope of this work, does not alter our conclusions. At the other end of the DTD, the earliest SD events obtained in this paper are a result of the mass stripping effect and are therefore dependent on its efficiency. Another way to obtain such early SD events is a WD + He star channel, as proposed by \citet{wang2009}. However, even without including this channel, our computations are assured of very early events (at 50 Myr) through the DD scenario.

The argument can also be extended to spiral galaxies, e.g. the Milky Way Galaxy. For the latter, it has been demonstrated in the past that the DD scenario can explain the Galactic birthrates well \citep[see e.g.][]{han1995,han1998}. The same has been done in previous studies with the Brussels code \citep[see][]{dedonder2004}, which is an additional indication that the difference between theoretical and observed rates in ellipticals may at least be partially caused by uncertainties on the observations in such galaxies.

\section{Conclusions}

We conclude that the single degenerate scenario alone cannot reproduce the observed distribution of delay times of type Ia supernovae. The double degenerate delay time distribution, possibly combined with the single degenerate one, does agree with the morphological shape of the observed distribution. However, assuming the $\alpha$-formalism for common envelope evolution, this is only true if Roche lobe overflow is treated to occur quasi-conservatively. In this case, most type Ia supernovae are formed through such a canonical Roche lobe overflow followed by a common envelope phase, not through two successive common envelope phases. The absolute number of events critically depends on a whole range of parameters and descriptions, a fact which may be used to gain further knowledge about them once more detailed observations of this absolute number become available. At this time however, the theoretical models are unable to reproduce the absolute number of observed type Ia supernovae, which is underestimated by a factor of at least three in elliptical galaxies with an age of 11 Gyr. This discrepancy is similar to those found in other population synthesis studies, but less pronounced in magnitude. It could be resolved by including the effects of rotation of binary components on their evolution.

\begin{acknowledgements}
      We thank Tomonori Totani for useful discussions regarding our comparison with observational absolute SN Ia rates, as well as Ashley Ruiter and the referee Zhanwen Han for constructive comments on the manuscript.
\end{acknowledgements}

\end{document}